\documentclass[reprint,superscriptaddress,amsmath,amssymb,pra,aps,showpacs]{revtex4-1}

\usepackage{amsmath,amsthm}
\usepackage {xcolor}
\usepackage{comment}

\usepackage{mathrsfs}
\usepackage{comment}
\usepackage{amssymb}
\usepackage{graphicx}
\usepackage{subfigure}

\usepackage[colorlinks,
            linkcolor=blue,
            anchorcolor=blue,
            citecolor=blue]{hyperref}
\newtheorem{definition}{Definition}
\newtheorem{theorem}{Theorem}

\newtheorem{corollary}{Corollary}

\usepackage{graphicx}% Include figure files
\usepackage{dcolumn}% Align table columns on decimal point
\usepackage{bm}% bold math
\begin{document}

%\preprint{APS/123-QED}

\title{Quantifying nonclassical correlations relative to local channels}

\author{Cong Xu}
\email[]{2230501028@cnu.edu.cn}
\affiliation{School of Mathematical Sciences, Capital Normal University, Beijing 100048, China}

\author{Tao Li}
\email[]{litao@btbu.edu.cn}
\affiliation{School of Mathematics and Statistics, Beijing Technology and Business University, Beijing 100048, China}

\author{Ruonan Ren}
\affiliation{School of Mathematics and Statistics, Shaanxi Normal University, Xi'an 710062,  China}

\author{Ming-Jing Zhao}
\email[]{zhaomingjingde@126.com}
\affiliation{School of Science, Beijing Information Science and Technology University, Beijing 102206, People's Republic of China}

\author{Shao-Ming Fei}
\email[]{feishm@cnu.edu.cn}
\affiliation{School of Mathematical Sciences, Capital Normal University, Beijing 100048, China}

\begin{abstract}
Nonclassical correlations are significant physical resources with extensive applications in quantum information processing. We introduce the modified Wigner-Yanase-Dyson skew information of a quantum state relative to a quantum channel, and a quantitative measure of quantum correlations. Their basic properties are explored in detail.
Through a specific example, we also compare our correlations measure with the existing one.  Moreover, the correlations relative to various channels including the von Neumann measurements, the unitary channels and the twirling channels are analyzed.


\end{abstract}

\maketitle

\section{Introduction}
A crucial issue in quantum information theory is to identify the valuable resources which outperform the corresponding classical protocols in quantum information tasks. Among these resources, quantum correlations are intrinsic and intriguing features of quantum mechanics and play an increasingly fundamental and important role in quantum information processing. \cite{Nielsen}.

Numerous approaches have been established to characterize various types of quantum correlations, such as entanglement detection through quantum Fisher information and Wigner-Yanase skew information \cite{fisher1,fisher2,fisher3}, as well as steerability identification based on uncertainty relations \cite{Sterring3,Sterring4,Sterring5}. Different characterizations of correlations from different perspectives have given rise to various definitions of correlations, such as quantum discord \cite{Discord1,Discord2,Discord3}, entanglement \cite{Entanglement1,Entanglement2,Entanglement3}, measurement-induced nonlocality \cite{non1,non2,non3}. As one of the most well-known correlations, entanglement has been quantified by a variety of methods, such as entanglement cost \cite{Entanglement4}, entanglement distillation \cite{Entanglement4}, entanglement of formation \cite{Entanglement5}, concurrence \cite{Entanglement5}. The quantum steering and Bell non-locality, which are quantum correlations stronger than entanglement, have also been widely studied \cite{Bell2,Sterring1,Sterring2}.

As different manifestations of the unique characteristics of quantum systems, there are distinctions as well as close connections between quantum coherence and quantum correlations. They both arise from the superposition principle and play an increasingly important role in quantum technology. Recently, much effort has been made to connect coherence
with correlations \cite{CC1,CC2,CC3,CC4,CC5,CC6,CC7,CC9,CC10,CC11}.

Since Baumgratz et al. \cite{BCP} proposed the framework for quantifying coherence, fruitful results have been obtained in characterizing quantum coherence both theoretically and experimentally \cite{Coherence1,Coherence3,Coherence4}. In addition, the concept of coherence relative to orthogonal bases has been extended to coherence relative to L\"{u}ders measurements (projection measurements), positive operator-valued measures (POVMs) and quantum channels \cite{Measures1,2018Luo,Measures3,Measures4,Measures5,Measures6}.

In 1963, Wigner and Yanase \cite{WY} introduced the concept of skew information, termed as Wigner-Yanase (WY) skew information. The WY skew information was later identified as the canonical version of quantum Fisher information and metric-adjusted skew information \cite{Fisher,Metric}. The skew information is often used to quantify quantum coherence \cite{2014kxiexinxi,2017yuchangshui}, quantum correlations \cite{2012fu,2013prl,2017epl,2018kim,2020fan} and quantum uncertainty \cite{2024xu1,2024xu2,2023zhang1,2021ren,2019fu,2023lihui}. Based on WY skew information, Luo et al. \cite{2012fu} proposed a correlations measure defined by the average difference of skew information. Following the ideas of \cite{2012fu}, Fan et al. \cite{2020fan} utilized the Wigner-Yanase-Dyson (WYD) skew information \cite{WY} to obtain similar results. In order to explore quantum correlations as resources in quantum metrology, Kim et al.\cite{2018kim} presented two methods for quantifying quantum correlations based on quantum Fisher information and demonstrated that when the quantum state is pure, these two quantifications are in consistent with the geometric discord.

In Ref. \cite{2018Luo}, the authors quantified coherence relative to channels in terms of WY skew information. By utilizing metric-adjusted skew information, Sun et al. \cite{2022Sun} extended the results of Ref. \cite{2018Luo}. Li et al. \cite{Li N} introduced a quantifier of correlations relative to a local channel as the coherence (based on WY skew information relative to channels) difference. Ren et al. \cite{Ren} generalized the results in Ref. \cite{Li N} to the metric-adjusted skew information version.

In this paper we introduce the concepts of modified Wigner-Yanase-Dyson (MWYD) skew information and MWYD skew information of a quantum state $\rho$ relative to a quantum channel $\Phi$ and investigate their fundamental properties. We then propose a quantifier of quantum correlations (relative to a local channel) as the difference between the global bipartite state $\rho_{AB}$ (relative to quantum channel $\Phi_A\otimes\mathcal{I}_B$) and the local quantum state $\rho_A$ (relative to a local quantum channel $\Phi_A$), and prove some basic properties satisfied by the quantifier.  We compare our correlations measure with the existing one by a detailed example. We then investigate the quantum correlations in bipartite states relative to some typical channels (measurements), such as the von Neumann measurements, the unitary channels and the twirling channels.

\section{Quantum correlations in terms of modified Wigner-Yanase-Dyson (MWYD) skew information}\label{sec2}

Let $H$ be a $d$-dimensional Hilbert space. Denote by $B(H)$, $S(H)$ and $D(H)$ the set of all bounded linear operators, Hermitian operators and density operators (positive operators with trace 1) on $H$, respectively. For a quantum state $\rho\in D(H)$ and an observable $A\in S(H)$, the Wigner-Yanase (WY) skew information \cite{WY} is defined by
\begin{equation*}
I(\rho,A):=-\frac{1}{2}\mathrm{tr}\left([\sqrt{\rho},A]^2\right).
\end{equation*}
 $I(\rho,A)$ is generalized by Dyson to
\begin{equation*}
\begin{aligned}
I_{\alpha}(\rho,A)
=&-\frac{1}{2}\mathrm{tr}([\rho^{\alpha},A][\rho^{1-\alpha},A])\\
=&\mathrm{tr}(\rho A^2)-\mathrm{tr}\rho^{\alpha}A\rho^{1-\alpha}A,\,\,\,\,\,\alpha\in(0,1),
\end{aligned}
\end{equation*}
which is now called the Wigner-Yanase-Dyson (WYD) skew
information \cite{WY}. By replacing observables with arbitrary operators $K\in B(H)$ (not necessarily Hermitian), Dou et.al \cite{Dou} proposed the generalized Wigner-Yanase-Dyson skew information,
\begin{equation}\label{AB}
\begin{aligned}
I_{\alpha}(\rho,K)
=&-\frac{1}{2}\mathrm{tr}([\rho^{\alpha},K][\rho^{1-\alpha},K^{\dag}])\\
=&\frac{1}{2}(\mathrm{tr}\rho KK^{\dag}-\mathrm{tr}\rho^{\alpha}K^{\dag}\rho^{1-\alpha}K+\mathrm{tr}\rho K^{\dag}K\\
-&\mathrm{tr}\rho^{\alpha}K \rho^{1-\alpha}K^{\dag}),\,\,\,\,\,\alpha\in(0,1).
\end{aligned}
\end{equation}
$I_{\alpha}(\rho,K)$ reduces to WYD skew information when $K\in S(H)$.

For a quantum state $\rho\in D(H)$ and an observable $A\in S(H)$, the variance of the observable $A$ in $\rho$ is defined by \cite{2004Luo,Gudder}
\begin{align}
 V(\rho,A)=\mathrm{tr}({\rho}A^2)-\mathrm{tr}({\rho}A)^2.
\end{align}
For arbitrary operators $K\in B(H)$, Gudder \cite{Gudder} proposed the generalized variance,
\begin{align}
 V(\rho,K)=\mathrm{tr}(\rho K^{\dag}K)-\mathrm{tr}({\rho}K)\mathrm{tr}({\rho}K^{\dag}).
\end{align}
$V(\rho,K)$ reduces to $V(\rho,A)$  when $K\in S(H)$. Observing the last two terms of equation (\ref{AB}), we find that when $\rho=|\psi\rangle\langle\psi|$ is a pure state, $\mathrm{tr}(|\psi\rangle\langle\psi|K^{\dag}K)-\mathrm{tr}(|\psi\rangle\langle\psi|K|\psi\rangle\langle\psi|K^{\dag})=V(|\psi\rangle\langle\psi|,K)$. Therefore, we provide the following definition.

\begin{definition}
We define the modified Wigner-Yanase-Dyson (MWYD) skew information as
\begin{equation}\label{eq4}
\begin{aligned}
T_{\alpha}(\rho,K)
=&\mathrm{tr}\rho K^{\dag}K-\mathrm{tr}\rho^{\alpha}K\rho^{1-\alpha}K^{\dag}, \,\,\,\alpha\in(0,1)
\end{aligned}
\end{equation}
for $K\in B(H)$.
\end{definition}

The quantities defined in (\ref{AB}) and (\ref{eq4}) are generally distinct for non-Hermitian operators. However, when the operator is Hermitian, both $I_{\alpha}(\rho,K)$ and $T_{\alpha}(\rho,K)$ reduce to the WYD skew information. As an extension of (\ref{eq4}), for a quantum channel $\Phi(\rho)=\sum_iK_i\rho K^{\dag}_i$ with Kraus
operators $\{K_i\}$ satisfying $\sum_{i}K_i^\dag K_i=\mathrm{I}$ (the identity operator) and a quantum state $\rho$, we give the following definition.

\begin{definition}\label{B}
We define the MWYD skew information of a quantum state $\rho$ relative to a quantum channel $\Phi$ as
\begin{equation}\label{STA}
\begin{aligned}
T_{\alpha}(\rho,\Phi)=&\sum_i T_{\alpha}(\rho,K_{i})\\
=&\sum_i(\mathrm{tr}\rho K^{\dag}_iK_i-\mathrm{tr}\rho^{\alpha}K_i\rho^{1-\alpha}K^{\dag}_i)\\
=&1-\mathrm{tr}\rho^{\alpha} \Phi (\rho^{1-\alpha}), \,\,\,\,\alpha\in(0,1).
\end{aligned}
\end{equation}
\end{definition}

\begin{theorem}\label{A}
The $T_{\alpha}(\rho,\Phi)$ has the following properties:

(i) (Non-negativity) $T_{\alpha}(\rho,\Phi)\geq0$ and $T_{\alpha}(\rho,\Phi)=0$ if $\Phi (\rho^{1-\alpha})=\rho^{1-\alpha}$.

(ii) (Ancillary independence)
\begin{equation*}
T_{\alpha}(\rho_{A},\Phi_{A})=T_{\alpha}(\rho_{A}\otimes\rho_{B},\Phi_{A}\otimes \mathcal{I}_B),
\end{equation*}
where $\rho_{A}$ and $\rho_{B}$ are quantum states of systems $A$ and
$B$, $\Phi_{A}$ and $\mathcal{I}_B$ are quantum channel and the identity channel on systems $A$ and $B$, respectively.

(iii) (Unitary covariance) For any unitary operator $U$,
\begin{equation}\label{U}
T_{\alpha}(U\rho\, U^{\dag},\Phi)=T_{\alpha}(\rho,\mathcal{U}^\dag\Phi\,\mathcal{U}),
\end{equation}
where $\mathcal{U}^\dag\Phi\,\mathcal{U}(\rho)=\sum_i(U^\dag K_{i}U)\rho(U^\dag K_{i}U)^{\dag}$
for $\Phi(\rho)=\sum_i{K_i\rho K^{\dag}_i}$.

(iv) (Linearity) $T_{\alpha}(\rho,\Phi)$ is positive-real-linear in the channel $\Phi$ in the sense that
\begin{equation*}
T_{\alpha}(\rho,\sum_i c_i\Phi_i)=\sum_i c_iT_{\alpha}(\rho,\Phi_i),
\end{equation*}
for any positive number $c_i$.

(v) (Decreasing under partial trace)
\begin{equation*}
T_{\alpha}(\rho_{AB},\Phi_{A}\otimes\mathcal{I}_B)\geq T_{\alpha}(\rho_{A},\Phi_{A}),
\end{equation*}
where $\rho_{AB}$ is any bipartite state on joint system $AB$.

(vi) (Convexity) $T_{\alpha}(\rho,\Phi)$ is convex with respect to $\rho$, i.e.,
\begin{equation*}
\sum_jp_jT_{\alpha}(\rho_j,\Phi)\geq T_{\alpha}(\sum_jp_j\rho_j,\Phi),
\end{equation*}
where $p_j\geq0$ for each $j$ with $\sum_j p_j=1$.
\end{theorem}

For the proof of Theorem 1, see Appendix A.  Although $T_{\alpha}(\rho,\Phi)$ satisfies some basic properties mentioned above, its physical significance is still unclear. Consequently, it is an interesting issue to consider the physical meaning of $T_{\alpha}(\rho,\Phi)$.  Based on the property (v) of $T_{\alpha}(\rho,\Phi)$, we introduce the following correlation measure.

\begin{definition}
We introduce the difference between the MWYD skew information of a global state $\rho_{AB}$  (relative to global channel $\Phi_A\otimes\mathcal{I}_B$) and the local state $\rho_A$ (relative to local channel $\Phi_A$) as a quantifier of correlation in $\rho_{AB}$ (relative to $\Phi_A$),
\begin{equation}\label{DT}
\begin{aligned}
D^T_{\alpha}(\rho_{AB}|\Phi_{A})
=T_{\alpha}(\rho_{AB},\Phi_{A}\otimes \mathcal{I}_B)-T_{\alpha}(\rho_{A},\Phi_{A}).
\end{aligned}
\end{equation}
\end{definition}

\begin{theorem}
The measure $D^T_{\alpha}(\rho_{AB}|\Phi_{A})$ of correlations has the following desirable properties, see proof in Appendix B:

(i) (Non-negativity) $D^T_{\alpha}(\rho_{AB}|\Phi_{A})\geq0$, with the equality holding when $\rho_{AB}=\rho_{A}\otimes\rho_{B}$ is a product state.

(ii) (Unitary covariance) For any local unitary operators $U_A$ and $U_B$ on systems $A$ and $B$, respectively, it holds that
\begin{equation*}
D^T_{\alpha}((U_A\otimes U_{B})\rho_{AB}(U_A\otimes U_{B})^\dag|\Phi_{A})=D^T_{\alpha}(\rho_{AB}|\mathcal{U}^\dag_A\Phi_{A}\mathcal{U}_A),
\end{equation*}
where $\mathcal{U}^\dag_A\Phi_{A}\,\mathcal{U}_A(\rho_A)=\sum_i(U^\dag_AK_{i}U_A)\rho_A(U^\dag_AK_{i}U_A)^{\dag}$
for $\Phi_{A}(\rho)=\sum{K_i\rho K^{\dag}_i}$.

(iii) (Contractivity) For any channel $\Phi_{B}$ on system $B$,
\begin{equation*}
D^T_{\alpha}(\mathcal{I}_A\otimes\Phi_{B}(\rho_{AB})|\Phi_{A})\leq D^T_{\alpha}(\rho_{AB}|\Phi_{A}),
\end{equation*}
where $\mathcal{I}_A$ denotes the identity channel on system $A$.
\end{theorem}

In \cite{Ren} the authors proposed the following correlation measures in terms of equation (\ref{AB}):
\begin{equation}\label{BB}
\begin{aligned}
D_{\alpha}(\rho_{AB}|\Phi_{A})
=I_{\alpha}(\rho_{AB},\Phi_{A}\otimes \mathcal{I}_B)-I_{\alpha}(\rho_{A},\Phi_{A}),
\end{aligned}
\end{equation}
where $I_{\alpha}(\rho,\Phi)=\sum_i I_{\alpha}(\rho,K_{i})$.
In the following example, we compare our correlations measure given by (\ref{DT}) with the one given by (\ref{BB}).

\noindent {\bf Example 1} Consider the following quantum state
$$
\rho_{AB}=\left(\begin{array}{cccc}
         \frac{1}{3}&\frac{1}{3}&-\frac{1}{3}&0\\
         \frac{1}{3}&\frac{1}{3}&-\frac{1}{3}&0\\
         -\frac{1}{3}&-\frac{1}{3}&\frac{1}{3}&0\\
         0&0&0&0\\
         \end{array}
         \right),
$$
and amplitude damping (AD) channel
$\Phi_{AD}(\rho)=\sum_{i=1}^2K_i\rho K_i^\dagger$
with the Kraus operators
$
K_1=
\begin{pmatrix}
1 & 0 \\
0 & \sqrt{1-p}
\end{pmatrix}
$,
$
K_2=
\begin{pmatrix}
0 & \sqrt{p} \\
0 & 0
\end{pmatrix}
$,
$0\leq p\leq1$.
By direct calculation, we have
\begin{equation}\label{TAB}
\begin{aligned}
&D^T_{\alpha}(\rho_{AB}|\Phi_{AD})\\
=&\frac{-1}{45}\times2^{-1-2\alpha}[-4^{1+\alpha}-6(3-\sqrt{5})^{2\alpha}(3+\sqrt{5})\\
&\times(\sqrt{1-p}-1)+6
(\sqrt{5}-3)(\sqrt{5}+3)^{2\alpha}(\sqrt{1-p}-1)\\
-&4^{1+\alpha}\sqrt{1-p}+p(3\times2^{1+2\alpha}-3(3-\sqrt{5})^{2\alpha}(1+\sqrt{5})\\
+&3(\sqrt{5}-1)(3+\sqrt{5})^{2\alpha})]
\end{aligned}
\end{equation}
and
\begin{equation}\label{IAB}
\begin{aligned}
&D_{\alpha}(\rho_{AB}|\Phi_{AD})\\
=&\frac{1}{45}\times4^{-1-\alpha}[2^{3+2\alpha}-12(3-\sqrt{5})^{2\alpha}(3+\sqrt{5})\\
&\times(\sqrt{1-p}-1)+12
(\sqrt{5}-3)(\sqrt{5}+3)^{2\alpha}(\sqrt{1-p}-1)\\
-&2^{3+2\alpha}\sqrt{1-p}+p(6\times2^{1+2\alpha}+3(3-\sqrt{5})^{2\alpha}(3+\sqrt{5})\\
-&3(\sqrt{5}-3)(3+\sqrt{5})^{2\alpha})].
\end{aligned}
\end{equation}
For $p=\frac{1}{4}$, Figure 1 shows that the correlations $D^T_{\alpha}(\rho_{AB}|\Phi_{AD})$
increases as the parameter $\alpha$ increases. In contrast, the correlations $D_{\alpha}(\rho_{AB}|\Phi_{AD})$ decreases when $\alpha$ increases for $\alpha\in (0,\frac{1}{2})$, while increases with $\alpha$ for $\alpha \in (\frac{1}{2}, 1)$. Moreover, we observe that $D_{\alpha}(\rho_{AB}|\Phi_{AD})$ is greater than $D^T_{\alpha}(\rho_{AB}|\Phi_{AD})$ for $\alpha\in (0,\frac{1}{2})$, whereas $D^T_{\alpha}(\rho_{AB}|\Phi_{AD})$ exceeds $D_{\alpha}(\rho_{AB}|\Phi_{AD})$ for $\alpha\in (\frac{1}{2},1)$, with both measures being equal at $\alpha=\frac{1}{2}$. For fixed $\alpha=\frac{3}{4}$, Figure 2 demonstrates that both correlations measures $D^T_{3/4}(\rho_{AB}|\Phi_{AD})$ and $D_{3/4}(\rho_{AB}|\Phi_{AD})$ increase as the parameter $p$ increases. Under this condition, our correlations measure $D^T_{3/4}(\rho_{AB}|\Phi_{AD})$ consistently remains greater than the correlations measure $D_{3/4}(\rho_{AB}|\Phi_{AD})$.
\begin{figure}[tbp]
  \centering
  \includegraphics[width=8.6cm]{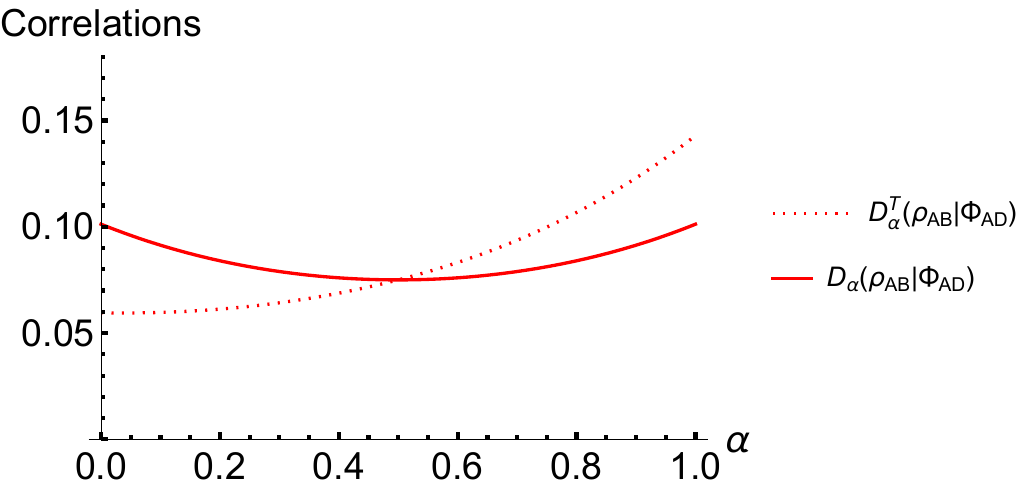}\\
\caption{$p=\frac{1}{4}$. The dotted red line and solid red line represent the the correlations $D^T_{\alpha}(\rho_{AB}|\Phi_{AD})$ and $D_{\alpha}(\rho_{AB}|\Phi_{AD})$, respectively, with respect to $\alpha$.} \label{fig:Fig1}
\end{figure}
\begin{figure}[tbp]
  \centering
  \includegraphics[width=8.6cm]{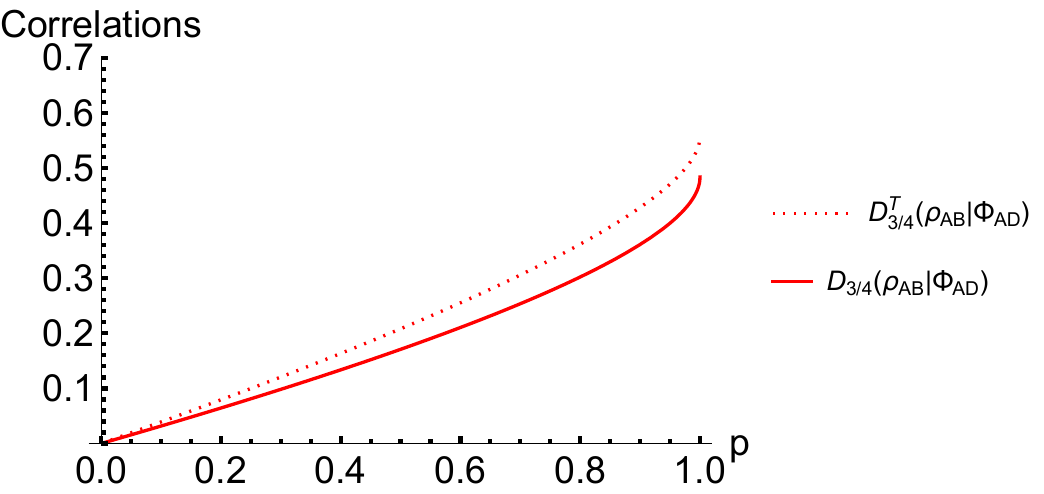}\\
\caption{$\alpha=\frac{3}{4}$. The dotted red line and solid red line represent the correlations $D^T_{3/4}(\rho_{AB}|\Phi_{AD})$ and $D_{3/4}(\rho_{AB}|\Phi_{AD})$, respectively, with respect to the parameter $p$.}\label{fig:Fig2}
\end{figure}

\section{APPLICATIONS}\label{sec3}
In this section, we illustrate the correlations $D^T_{\alpha}(\rho_{AB}|\Phi_{A})$ by studying the correlations of some special channels such as von Neumann measurements, unitary channels and twirling channels.

\subsection{Correlations relative to a von Neumann measurement}

Let $\Pi=\{\Pi_i=|i\rangle\langle i|:i=1,2,\ldots,d_A\}$ be a von Neumann measurement on system $A$. Taking the quantum channel $\Phi$ as a von Neumann measurement $\Pi$, we have
\begin{equation}\label{VOSTA}
\begin{aligned}
T_{\alpha}(\rho,\Pi)=&1-\sum_i\langle i|\rho^{1-\alpha}|i\rangle\langle i|\rho^{\alpha}|i\rangle\\
=&I_{\alpha}(\rho,\Pi).
\end{aligned}
\end{equation}

$I_{\alpha}(\rho,\Pi)$ has been proven to be a well-defined coherence measure \cite{Fan,2022Sun}.
In particular, taking $\Pi_A\otimes \mathcal{I}_B=\{\Pi^i_A\otimes \mathrm{I}_B\}$, where $\Pi_A=\{\Pi^i_A\}$ is the von Neumann measurement on system $A$, we have
\begin{equation}\label{min}
\begin{aligned}
G(\rho_{AB})=\min \limits_{\Pi_{A}}T_{\frac{1}{2}}(\rho_{AB},\Pi_{A}\otimes \mathcal{I}_B),
\end{aligned}
\end{equation}
where $G(\rho_{AB})$ is just the geometric discord given in  \cite{2017epl}. Here $\mathcal{I}_B$ and $\mathrm{I}_B$ are the identity channel and the identity operator on systems $B$, respectively.

According to the property (v) of $T_{\alpha}(\rho,\Phi)$, we see that the coherence of a global state $\rho_{AB}$ (relative to $\Pi_A\otimes\mathcal{I}_B$) is greater than or equal to that of the local state $\rho_A$ (relative to $\Pi_A$).

Correspondingly, we introduce the following quantity to quantify correlations based on the difference of coherence:
\begin{equation}\label{VODT}
\begin{aligned}
&D^T_{\alpha}(\rho_{AB}|\Pi_{A})\\
\equiv&T_{\alpha}(\rho_{AB},\Pi_{A}\otimes \mathcal{I}_B)-T_{\alpha}(\rho_{A},\Pi_{A})\\
=&T_{\alpha}(\rho_{AB},\Pi_{A}\otimes\mathcal{I}_B)
-T_{\alpha}(\rho_{A}\otimes\rho_{B},\Pi_{A}\otimes \mathcal{I}_B).
\end{aligned}
\end{equation}
In fact, $D^T_{\alpha}(\rho_{AB}|\Pi_{A})$ can also be reinterpreted as the coherence difference between $\rho_{AB}$ and the corresponding product state $\rho_{A}\otimes\rho_{B}$ relative to the same product channel $\Pi_{A}\otimes \mathcal{I}_B$. Obviously, $D^T_{\alpha}(\rho_{AB}|\Pi_{A})$ satisfies all the properties of Theorem 2. It is a special case of the correlations measure given in Ref. [60]. To eliminate the dependence of $D^T_{\alpha}(\rho_{AB}|\Pi_{A})$ on the von Neumann measurements, and capture some intrinsic correlations in $\rho_{AB}$, we take the minimum and maximum over the von Neumann measurement $\Pi_A=\{\Pi^i_A\}$ on system $A$ and define the following correlations measures:

(a) Maximal coherence difference
\begin{equation}
\widehat{D^T_{\alpha}}(\rho_{AB})=\max \limits_{\Pi_A}D^T_{\alpha}(\rho_{AB}|\Pi_{A}),
\end{equation}\\

(b) Minimal coherence difference
\begin{equation}
\overline{D^T_{\alpha}}(\rho_{AB})=\min \limits_{\Pi_A}D^T_{\alpha}(\rho_{AB}|\Pi_{A}),
\end{equation}\\

$\widehat{D^T_{\alpha}}(\rho_{AB})$ and $\overline{D^T_{\alpha}}(\rho_{AB})$, defined respectively as the maximum and minimum coherence difference, characterize some different aspects of quantum states. They have the similar properties of the following correlations measure \cite{2012fu},
\begin{equation}
\begin{aligned}
D(\rho^{AB})=\sum^{d^2}_{l=1}(I(\rho_{AB},W^l_{A}\otimes \mathcal{I}_B)-I(\rho_{A},W^i_{A})),
\end{aligned}
\end{equation}
where $\{W^l_A\}$ is an orthonormal base for the real Hilbert space $L(H)$ of all observables on $H$, with the Hilbert-Schmidt inner product $\langle A,B\rangle=\mathrm{tr}AB$.

\begin{corollary}
The quantity $\widehat{D^T_{\alpha}}(\rho_{AB})$ also has the following properties, see Appendix C:

(i) (Non-negativity) $\widehat{D^T_{\alpha}}(\rho_{AB})\geq0$, with equality holding if and only if  $\rho_{AB}=\rho_{A}\otimes\rho_{B}$ is a product state.

(ii) (Unitary covariance) For any local unitary operators $V_A$ and $V_B$ on systems $A$ and $B$, respectively, it holds that
\begin{equation*}
\widehat{D^T_{\alpha}}((V_A\otimes V_{B})\rho_{AB}(V_A\otimes V_{B})^\dag)=\widehat{D^T_{\alpha}}(\rho_{AB}).
\end{equation*}

(iii) (Contractivity) For any channel $\Phi_{B}$ on system $B$,
\begin{equation*}
\widehat{D^T_{\alpha}}(\mathcal{I}_A\otimes\Phi_{B}(\rho_{AB}))\leq \widehat{D^T_{\alpha}}(\rho_{AB}).
\end{equation*}
\end{corollary}

Note that when $\alpha=\frac{1}{2}$ and $\Pi_{A}$ does not disturb the local state $\rho_A$, i.e., $\Pi_{A}(\rho_A)=\rho_A$, one has
\begin{equation}\label{max}
\begin{aligned}
\widehat{D^T_{\frac{1}{2}}}(\rho_{AB})=\max \limits_{\Pi_A}T_{\frac{1}{2}}(\rho_{AB},\Pi_{A}\otimes \mathcal{I}_B),
\end{aligned}
\end{equation}
which coincides with the measurement-induced nonlocality introduced in Ref.\cite{2016Li}.

So far we have proven that the amount of the maximum correlations in
$\rho_{AB}$ relative to the local measurement $\Pi_A$ vanishes if
and only if $\rho_{AB}$ has vanishing correlations. This
provides a characterization of product states in terms of local von Neumann measurement on system $A$. In the following Corollary 2, we show that by taking the minimum correlations, $\overline{D^T_{\alpha}}(\rho_{AB})$ can be applied to characterize some natural states without quantum discord.

\begin{corollary}
The quantity $\overline{D^T_{\alpha}}(\rho_{AB})$ has the following properties, see Appendix C:

(i) (Non-negativity) $\overline{D^T_{\alpha}}(\rho_{AB})\geq0$, with the equality holding when $\rho_{AB}$ is a classical-quantum state.

(ii) (Unitary covariance) For any local unitary operators $V_A$ and $V_B$ on systems $A$ and $B$, respectively, it holds that
\begin{equation*}
\overline{D^T_{\alpha}}((V_A\otimes V_{B})\rho_{AB}(V_A\otimes V_{B})^\dag)=\overline{D^T_{\alpha}}(\rho_{AB}).
\end{equation*}

(iii) (Contractivity) For any channel $\Phi_{B}$ on system $B$,
\begin{equation*}
\overline{D^T_{\alpha}}(\mathcal{I}_A\otimes\Phi_{B}(\rho_{AB}))\leq \overline{D^T_{\alpha}}(\rho_{AB}).
\end{equation*}
\end{corollary}

\subsection{Correlations relative to a unitary channel}

Unitary channels are frequently utilized and hold significant importance in the fields of quantum information theory and quantum computation \cite{Nielsen}. For a local unitary channel $\mathcal{U}$ on system $A$, we have $\mathcal{U}(\rho)=U\rho U^{\dag}$ for any state $\rho$ on system $A$. When we take the quantum channel $\Phi$ as a unitary channel $\mathcal{U}$, $T_{\alpha}(\rho,\Phi)$ reduces to:
\begin{equation}
T_{\alpha}(\rho,\mathcal{U})=1-\mathrm{tr}\rho^{\alpha} \mathcal{U} (\rho^{1-\alpha}), \,\,\,\,\alpha\in(0,1).
\end{equation}

Correspondingly, we define a quantity to quantify correlations in $\rho_{AB}$ relative to the local unitary channel $\mathcal{U}_A$,
\begin{equation}\label{mmm}
\begin{aligned}
&D^T_{\alpha}(\rho_{AB}|\mathcal{U}_{A})\\
\equiv&T_{\alpha}(\rho_{AB},\mathcal{U}_{A}\otimes \mathcal{I}_B)-T_{\alpha}(\rho_{A},\mathcal{U}_{A})\\
=&T_{\alpha}(\rho_{AB},\mathcal{U}_{A}\otimes\mathcal{I}_B)-T_{\alpha}(\rho_{A}\otimes\rho_{B},\mathcal{U}_{A}\otimes \mathcal{I}_B).
\end{aligned}
\end{equation}

To eliminate the reliance on unitary operators, we take minimization, maximization or integration over all the unitary operators on system $A$.
Obviously, $\min\limits_{{U}_A}D^T_{\alpha}(\rho_{AB}|\mathcal{U}_{A})
=T_{\alpha}(\rho_{AB},\mathcal{I}_A\otimes \mathcal{I}_B)-T_{\alpha}(\rho_{A},\mathcal{I}_A)=0$. Taking the maximization, we define
\begin{equation}
\widetilde{D^T_{\alpha}}(\rho_{AB})=\max \limits_{U_A}D^T_{\alpha}(\rho_{AB}|\mathcal{U}_{A}).
\end{equation}

\begin{corollary}\label{D}
The correlation measure $\widetilde{D^T_{\alpha}}(\rho_{AB})$ has the following desirable properties, see the proof in Appendix C:

(i) (Non-negativity) $\widetilde{D^T_{\alpha}}(\rho_{AB})\geq0$, and the equality holds if $\rho_{AB}=\rho_{A}\otimes\rho_{B}$ is a product state.

(ii) (Unitary covariance) For any local unitary operators $V_A$ and $V_B$ on systems $A$ and $B$, respectively, it holds that
\begin{equation*}
\widetilde{D^T_{\alpha}}((V_A\otimes V_{B})\rho_{AB}(V_A\otimes V_{B})^\dag)=\widetilde{D^T_{\alpha}}(\rho_{AB}).
\end{equation*}

(iii) (Contractivity) For any channel $\Phi_{B}$ on system $B$,
\begin{equation*}
\widetilde{D^T_{\alpha}}(\mathcal{I}_A\otimes\Phi_{B}(\rho_{AB}))\leq \widetilde{D^T_{\alpha}}(\rho_{AB}).
\end{equation*}
\end{corollary}

\subsection{Correlations relative to twirling channel induced by unitary group}

Let $U(H_A)$ be a unitary group on a $d$-dimensional system $A$. The twirling channel $\mathcal{T}_{U(H_A)}$ is defined by
\begin{equation}
\mathcal{T}_{U(H_A)}(\rho_{AB})=\int_{U(H_A)}U_A\rho_{AB} U^{\dag}_A dU_A,
\end{equation}
where $dU_A$ is the normalized Haar measure on $U(H_A)$. The MWYD skew information of $\rho_{AB}$ (relative to channel $U(H_A)\otimes\mathcal{I}_B$) and the local state $\rho_A=\mathrm{tr}_B\rho_{AB}$ (relative to local channel $U(H_A)$) are defined as
\begin{equation}
T_{\alpha}(\rho_A,\mathcal{T}_{U(H_A)})=\int_{U(H_A)} T_{\alpha}(\rho_A,U_A) dU_A,
\end{equation}
 and
\begin{equation}
T_{\alpha}(\rho_{AB},\mathcal{T}_{U(H_A)}\otimes\mathcal{I}_B)=\int_{U(H_A)} T_{\alpha}(\rho_{AB},U_A\otimes\mathrm{I}_B) dU_A,
\end{equation}
respectively.  The corresponding quantifier of correlations relative to $\mathcal{T}_{U(H_A)}$ is given by
\begin{equation}
\begin{aligned}
&D^T_{\alpha}(\rho_{AB}|\mathcal{T}_{U(H_A)})\\
=&\int_{U(H_A)}T_{\alpha}(\rho_{AB},U_A\otimes\mathrm{I}_B)-T_{\alpha}(\rho_A,U_A) dU_A.
\end{aligned}
\end{equation}

By direct calculation, we obtain
\begin{equation}
\begin{aligned}
&D^T_{\alpha}(\rho_{AB}|\mathcal{T}_{U(H_A)})
=\int_{U(H_A)}\mathrm{tr}(\rho^{\alpha}_AU_A\rho^{1-\alpha}_AU^{\dag}_A)   dU_A\\
-&\int_{U(H_A)}\mathrm{tr}[\rho^{\alpha}_{AB}(U_A\otimes\mathrm{I}_B)\rho^{1-\alpha}_{AB}(U^{\dag}_A\otimes\mathrm{I}_B)]   dU_A\\
=&\frac{1}{d}(\mathrm{tr}\rho^{\alpha}_A\mathrm{tr}\rho^{1-\alpha}_A)-\mathrm{tr}\left(\rho^{\alpha}_{AB}\left(\frac{\mathrm{I}_A}{d}\otimes\mathrm{tr}_A(\rho^{1-\alpha}_{AB})\right)\right)\\
=&\frac{1}{d}[\mathrm{tr}\rho^{\alpha}_A\mathrm{tr}\rho^{1-\alpha}_A-\mathrm{tr}_B(\mathrm{tr}_A(\rho^{\alpha}_{AB})\mathrm{tr}_A(\rho^{1-\alpha}_{AB}))],
\end{aligned}
\end{equation}
where the second equality comes from the following identities \cite{Zhang},
\begin{equation}
\begin{aligned}
\int_{U(H_A)}U_A X U^{\dag}_A dU_A=\mathrm{tr}X\frac{\mathrm{I}_A}{d}
\end{aligned}
\end{equation}
for any operator $X$ on system $A$, and
\begin{equation}
\begin{aligned}
\int_{U(H_A)}(U_A\otimes\mathrm{I}_B)T_{AB}(U^{\dag}_A\otimes\mathrm{I}_B) dU_A=\frac{\mathrm{I}_A}{d}\otimes\mathrm{tr}_{A}T_{AB}
\end{aligned}
\end{equation}
for any operator $T_{AB}$ on the joint system $AB$. Here $\mathrm{I}_A$ is the identity operator on system $A$.

Let $\{W_l: l=1, 2,\ldots, d^2\}$ be the orthonormal basis for all Hermite operators on the real Hilbert space $L(H)$ with the Hilbert Schmidt inner product $\langle A,B\rangle=\mathrm{tr}AB$ and $\sum^{d^2}_{l=1}
W^2_{l}=d\mathrm{I}_A$. For the completely depolarizing channel on system $A$,
\begin{equation}
\begin{aligned}
\Lambda_A(\rho)=\frac{1}{d}\sum^{d^2}_{l=1}
W_l\rho W_l,
\end{aligned}
\end{equation}
with the Kraus operators $\{\frac{W_l}{\sqrt{d}}:l=1,2,\ldots,d^2\}$, we have the following theorem, see the proof in Appendix D.

{\bf Theorem 3} The correlation of any bipartite state $\rho_{AB}$ relative to the twirling
channel $\mathcal{T}_{U(H_A)}$ is consistent with the completely depolarizing channel $\Lambda_A$,
\begin{equation}
\begin{aligned}
&D^T_{\alpha}(\rho_{AB}|\mathcal{T}_{U(H_A)})
=D^T_{\alpha}(\rho_{AB}|\Lambda_A).
\end{aligned}
\end{equation}

\section{Conclusion}\label{sec4}
We have introduced the new definitions of MWYD skew information of a local state $\rho$ relative to local channel $\Phi$ and explored their elegant properties. Furthermore, we have quantified quantum correlations as the difference between the MWYD skew information of a global state $\rho_{AB}$ (relative to global channel $\Phi_A\otimes\mathcal{I}_B$) and the local state $\rho_A$ (relative to local channel $\Phi_A$).  Additionally, we have compared our correlations measure with the existing one and provided explicit expressions for the correlations of a quantum state, with respect to amplitude damping channel, as well as graphically depicted the dynamic changes in correlations. Finally, we have presented the correlations relative to von Neumann measurement, unitary channel and twirling channel.

Although the quantity $T_{\alpha}(\rho,K)$ coincides with WYD skew information for Hermitian operators, its operational significance for non-Hermitian operators remains unclear. Exploring the physical meanings of $T_{\alpha}(\rho,K)$ is also a very interesting question. In Ref. \cite{2018Luo},
the authors introduced $I(\rho,\Phi)$ as an extension of the WY skew information and provided its physical interpretation: asymmetry, coherence, non-commutativity, quantumness, and quantum uncertainty. Furthermore, the complementary relationship between symmetry and asymmetry provided by $I(\rho,\Phi)$ can be interpreted as a quantification of Bohr's complementarity principle. Correspondingly, exploring the physical applications of our newly proposed $T_\alpha(\rho,\Phi)$ is also a very interesting question.

\section{APPENDIX}\label{sec7}
Here we present the detailed proofs of Theorems 1-3 and Corollaries 1-3.

\subsection{Proof of Theorem 1}

\begin {proof}
 $T_{\alpha}(\rho,\Phi)\geq 0$ is equivalent to $\mathrm{tr}\rho^{\alpha} \Phi (\rho^{1-\alpha})\in[0,1]$.  Because $\rho^{\frac{\alpha}{2}} K_i\rho^{1-\alpha} K^{\dag}_i \rho^{\frac{\alpha}{2}}$ is a positive matrix, one has
$\mathrm{tr}\rho^{\alpha} \Phi (\rho^{1-\alpha})=\sum_i\mathrm{tr}\rho^{\alpha} K_i\rho^{1-\alpha}K^{\dag}_i\geq0$. Suppose that $F(\alpha)=\sum_i(\mathrm{tr}\rho^{\alpha}K_i\rho^{1-\alpha}K^{\dag}_i)
=\sum_{imn}\lambda^{\alpha}_m\lambda^{1-\alpha}_n|\langle m|K_i|n\rangle|^2$, where the second equality follows from  spectral decomposition. By taking the derivative of $\alpha$, we obtain $F'(\alpha)=\sum_{imn}\lambda^{\alpha}_m\lambda^{1-\alpha}_n(\ln\lambda_m-\ln\lambda_n)|\langle m|K_i|n\rangle|^2$ and $F''(\alpha)=\sum_{imn}\lambda^{\alpha}_m\lambda^{1-\alpha}_n|(\ln\lambda_m-\ln\lambda_n)|^2|\langle m|K_i|n\rangle|^2\geq0$, i.e., $F(\alpha)$ is convex with respect to $\alpha$. By simply calculation, we get $F(0)=\sum_i\mathrm{tr}K_i\rho K^{\dag}_i=1$ and  $F(1)=\sum_i\mathrm{tr}\rho K_i K^{\dag}_i\leq1$,
therefore, $\mathrm{tr}\rho^{\alpha} \Phi (\rho^{1-\alpha})=F(\alpha)\leq1$.
It is obvious that if $\Phi (\rho^{1-\alpha})=\rho^{1-\alpha}$, $T_{\alpha}(\rho,\Phi)=0$. Therefore, item (i) is proved.

Items (ii)-(iii) are clear by straightforward manipulation of the definition.

For item (iv), suppose $\Phi_1(\rho)=\sum_{i=1}^fK_i\rho K^{\dag}_i$, $\Phi_2(\rho)=\sum_{j=1}^mL_j\rho L^{\dag}_j$ and $\widetilde{\Phi}(\rho)=\sum_{s=1}^{f+m}F_s\rho F^{\dag}_s$, where
$$ F_s=\left\{
\begin{aligned}
K_s, &&&&s \in\{1,2,\cdots, f\}\\
L_{s-f}, &&&&s \in\{f+1,f+2,\cdots, f+m\}
\end{aligned}
\right.
$$
i.e., $\widetilde{\Phi}(\rho)=\Phi_1(\rho)+\Phi_2(\rho)$. Suppose $\{K_i\}$ is a set of the Kraus operators of quantum channel $\Phi$. Then $\{\sqrt{c}K_i\}$ is the set of the Kraus operators of $c\Phi$, where $c>0$. For $c_1,c_2>0$ and quantum channels $\Phi_1$, $\Phi_2$, direct calculation we have
\begin{equation*}
T_{\alpha}(\rho, c_1\Phi_1+c_2\Phi_2)=c_1T_{\alpha}(\rho,\Phi_1)+c_2T_{\alpha}(\rho,\Phi_2).
\end{equation*}
It is not hard to prove item (iv) by induction.

By simply calculation, we obtain the following inequality
\begin{equation*}\label{STAB-STA}
\begin{aligned}
&T_{\alpha}(\rho_{AB},\Phi_{A}\otimes\mathcal{I}_B)-T_{\alpha}(\rho_{A},\Phi_{A})\\
=&[\mathrm{tr}\rho^{\alpha}_{A} \Phi_A
(\rho^{1-\alpha}_{A})-\mathrm{tr}\rho^{\alpha}_{AB} \Phi_A\otimes \mathcal{I}_B (\rho^{1-\alpha}_{AB})]\\
=&\sum_i[\mathrm{tr}\rho^{\alpha}_{A} K_i\rho^{1-\alpha}_{A}K^{\dag}_i-\mathrm{tr}\rho^{\alpha}_{AB}(K_i\otimes \mathrm{I}_{B})\rho^{1-\alpha}_{AB}(K_i\otimes \mathrm{I}_{B})^\dag]\\
\geq&0,
\end{aligned}
\end{equation*}
where the last inequality follows from the Corollary 1.3 in \cite{Lieb}. Hence, item (v) holds.

Let $Y$ is an arbitrary operator, and $0\leq t\leq1$. By using
the Lieb's theorem \cite{Nielsen}, the function
$f(A,B)=\mathrm{tr}(Y^\dagger A^t YB^{1-t})$ is jointly concave in
positive matrices $A$ and $B$, i.e.,
 $f\left(\sum_jp_jA_j,\sum_jp_jB_j\right)\geq\sum_jp_jf(A_j,B_j)$, where $p_j\geq 0$ with $\sum_jp_j=1$, and $A_j$, $B_j$ be positive operators for each $j$.
Taking $A_j=B_j=\rho_j, t=\alpha$, and $Y=K_i$, one obtains
$\mathrm{tr}\left[(\sum_jp_j\rho_j)^{\alpha}K_i(\sum_jp_j\rho_j)^{1-\alpha}K_i^\dag\right]\geq\sum_jp_j\mathrm{tr}\rho_j^{\alpha}K_i\rho_j^{1-\alpha}K_i^\dag$
for each $i$. Item (vi) follows immediately by summing over $i$ on both sides of the two inequalities.
\end {proof}

\subsection{Proof of Theorem 2}

\begin {proof} (i) From the properties (ii) and (v) of $T_{\alpha}(\rho,\Phi)$, it is obvious that $D^T_{\alpha}(\rho_{AB}|\Phi_{A})\geq0$, the equality holds when $\rho_{AB}=\rho_{A}\otimes\rho_{B}$ is a product state.

(ii) According to equations $(\ref{U})$ and $(\ref{DT})$, we have
\begin{equation*}
\begin{aligned}
&D^T_{\alpha}((U_A\otimes U_{B})\rho_{AB}(U_A\otimes U_{B})^\dag|\Phi_{A})\\
=&T_{\alpha}((U_A\otimes U_{B})\rho_{AB}(U_A\otimes U_{B})^\dag,\Phi_{A}\otimes \mathcal{I}_B)\\
&-T_{\alpha}((U_A\otimes U_{B})(\rho_{A}\otimes\rho_{B})(U_A\otimes U_{B})^\dag,\Phi_{A}\otimes \mathcal{I}_B)\\
=&T_{\alpha}(\rho_{AB},\mathcal{U}^\dag_A\Phi_{A}\mathcal{U}_A\otimes \mathcal{I}_B)-T_{\alpha}(\rho_{A}\otimes\rho_{B},\mathcal{U}^\dag_A\Phi_{A}\mathcal{U}_A\otimes \mathcal{I}_B)\\
=&D^T_{\alpha}(\rho_{AB}|\mathcal{U}^\dag_A\Phi_{A}\mathcal{U}_A).
\end{aligned}
\end{equation*}

(iii) For any channel $\Phi_B$ on system $B$, we can always choose an ancillary system $C$, a state $\rho_C$, and a unitary operator $U_{BC}$ on $BC$, such that $\Phi_B(\rho_B) = \mathrm{tr}_CU_{BC}(\rho_B\otimes \rho_C)U_{BC}^{\dag}$. According to the property (ii) and (v) of $T_{\alpha}(\rho,\Phi)$, we get
\begin{equation}\label{ABC}
\begin{aligned}
&T_{\alpha}(\mathcal{I}_A\otimes\Phi_{B}(\rho_{AB}),\Phi_{A}\otimes\mathcal{I}_B)\\
=&T_{\alpha}(\mathrm{tr}_C(\mathrm{I}_A\otimes U_{BC})(\rho_{AB}\otimes \rho_C)( \mathrm{I}_A\otimes U_{BC})^{\dag},\Phi_{A}\otimes\mathcal{I}_B)\\
\leq&T_{\alpha}((\mathrm{I}_A\otimes U_{BC})(\rho_{AB}\otimes \rho_C)( \mathrm{I}_A\otimes U_{BC})^{\dag},\Phi_{A}\otimes\mathcal{I}_B\otimes\mathcal{I}_C)\\
=&T_{\alpha}(\rho_{AB}\otimes \rho_C,(\mathcal{I}_A\otimes \mathcal{U}_{BC})^{\dag}(\Phi_{A}\otimes\mathcal{I}_B\otimes\mathcal{I}_C)(\mathcal{I}_A\otimes \mathcal{U}_{BC}))\\
=&T_{\alpha}(\rho_{AB}\otimes \rho_C,\Phi_{A}\otimes\mathcal{I}_B\otimes\mathcal{I}_C)\\
=&T_{\alpha}(\rho_{AB},\Phi_{A}\otimes\mathcal{I}_B).
\end{aligned}
\end{equation}

Consequently,
\begin{equation*}
\begin{aligned}
&D^T_{\alpha}(\mathcal{I}_A\otimes\Phi_{B}(\rho_{AB})|\Phi_{A})\\
=&T_{\alpha}\left(\mathcal{I}_A\otimes\Phi_{B}\left(\rho_{AB}\right),\Phi_A\otimes\mathcal{I}_B\right)\\
-&T_{\alpha}\left(\mathcal{I}_A\otimes\Phi_{B}\left(\rho_A\otimes\rho_B\right),\Phi_A\otimes\mathcal{I}_B\right)\\
=&T_{\alpha}\left(\mathcal{I}_A\otimes\Phi_{B}\left(\rho_{AB}\right),\Phi_A\otimes\mathcal{I}_B\right)
-T_{\alpha}\left(\rho_A,\Phi_A\right)\\
\leq&T_{\alpha}\left(\rho_{AB},\Phi_A\otimes\mathcal{I}_B\right)
-T_{\alpha}\left(\rho_A,\Phi_A\right)\\
=& D^T_{\alpha}\left(\rho_{AB}|\Phi_A\right).
\end{aligned}
\end{equation*}
\end {proof}

\subsection{Proof of Corollaries 1-3}

\begin {proof}
A bipartite state $\rho_{AB}$ has vanishing quantum discord if it can be written as
\begin{equation*}
\begin{aligned}
\rho_{AB}=\sum^{d_A}_jp_j|j\rangle\langle j|\otimes\rho^B_j,
\end{aligned}
\end{equation*}
where $p_j$ is a probability distribution, $\{|j\rangle:j=1,2,\ldots,d_A\}$ is an  orthonormal basis of the system Hilbert space $H_A$,
and $\{\rho^B_j\}\in H_B$. The reduced state $\rho_A=\mathrm{tr}_B\rho_{AB}=\sum_jp_j|j\rangle\langle j|$. In this case, the $\rho_{AB}$ is also called a classical-quantum correlated state \cite{2008 Luo}.

Since the minimum/maximum is taken over all von Neumann measurements, it is evident that item (i) of Corollary 1-2 follows from equation (\ref{VODT}). According to equation (\ref{mmm}), $\widetilde{D^T_{\alpha}}(\rho_{AB})\geq0$ and the equality holds if $\rho_{AB}=\rho_{A}\otimes\rho_{B}$ is a product state. For items (ii) and (iii) of Corollaries 1-3, the proof process is similar to that of Theorem 2. As product states are a subset of classical quantum states \cite{2024Li}, we have
$\widehat{D^T_{\alpha}}(\rho_{AB})=\overline{D^T_{\alpha}}(\rho_{AB})=\widetilde{D^T_{\alpha}}(\rho_{AB})=0$ if $\rho_{AB}=\rho_A\otimes\rho_B$.
\end {proof}

{\subsection{Proof of Theorem 3}
\begin {proof} By utilizing the following facts \cite{Fanya}:
\begin{equation}
\begin{aligned}
\sum^{d^2}_{l=1}(W_l\otimes\mathrm{I}_B)T_{AB}(W_l\otimes\mathrm{I}_B)=\mathrm{I}_A\otimes \mathrm{tr}_{A}T_{AB},
\end{aligned}
\end{equation}
for any bipartite state $T_{AB}$ and $\sum^{d^2}_{l=1}
W^2_l=d\mathrm{I}_A$.
We have
\begin{equation}
\begin{aligned}
&D^T_{\alpha}(\rho_{AB}|\Lambda_A)\\
=&\sum^{d^2}_{l=1}T_{\alpha}\left(\rho_{AB},\frac{W_l}{\sqrt{d}}\otimes\mathrm{I}_B\right)-\sum^{d^2}_{l=1}T_{\alpha}\left(\rho_{A},\frac{W_l}{\sqrt{d}}\right)   \\
=&\frac{1}{d}\sum^{d^2}_{l=1}[T_{\alpha}\left(\rho_{AB},W_l\otimes\mathrm{I}_B\right)-\sum^{d^2}_{l=1}T_{\alpha}\left(\rho_{A},W_l\right)]\\
=&\frac{1}{d}\sum^{d^2}_{l=1}[\mathrm{tr}\rho^{\alpha}_A W_l\rho^{1-\alpha}_A W_l-\mathrm{tr}\rho^{\alpha}_{AB} (W_l\otimes\mathrm{I}_B)\rho^{1-\alpha}_{AB}(W_l\otimes\mathrm{I}_B)]\\
=&\frac{1}{d}[\mathrm{tr}\rho^{\alpha}_A\mathrm{tr}\rho^{1-\alpha}_A-\mathrm{tr}(\rho^{\alpha}_{AB}(\mathrm{I}_A\otimes \mathrm{tr}_A\rho^{1-\alpha}_{AB}))]\\
=&\frac{1}{d}[\mathrm{tr}\rho^{\alpha}_A\mathrm{tr}\rho^{1-\alpha}_A-\mathrm{tr}_B(\mathrm{tr}_A(\rho^{\alpha}_{AB})\mathrm{tr}_A(\rho^{1-\alpha}_{AB}))],
\end{aligned}
\end{equation}
which completes the proof.
\end {proof}

\bigskip
\noindent{\bf Acknowledgments}\, \,
M.J.Zhao thanks the center for Quantum Information, Institute for Interdisciplinary Information Sciences, Tsinghua University for hospitality. This work was supported by National Natural Science Foundation of China (Grant Nos. 12171044,12175147); the specific research fund of the Innovation Platform for Academicians of Hainan Province; the Disciplinary funding of Beijing Technology and Business University; the Fundamental Research Funds For the Central Universities (Grant No: LHRCTS23057).

\bigskip
\bibliographystyle{apsrev4-2}

\end{document}